\documentclass[twocolumn]{aastex631}

\usepackage[utf8]{inputenc}
\usepackage{textgreek}

\begin{document}

\title{Distillation of $^{56}$Fe in Ultramassive O-Ne White Dwarfs}

\author{Matthew E. Caplan}
\affiliation{Department of Physics, Illinois State University \\ Normal, IL 67190, USA}

\author{Simon Blouin}
\affiliation{Department of Physics and Astronomy, University of Victoria \\ Victoria, BC V8W 2Y2, Canada}

\author{Ian F. Freeman}
\affiliation{Department of Physics, Illinois State University \\ Normal, IL 67190, USA}



\begin{abstract}
When white dwarfs freeze the plasma mixtures inside them undergo separation processes which can produce radical changes in the composition profile of the star. The abundance of neutron rich elements, such as $^{22}$Ne or $^{56}$Fe, determines whether or not the first crystals are more or less dense than the surrounding fluid and thus whether they sink or float. These processes have now been studied for C-O-Ne and C-O-Fe mixtures, finding that distillation and precipitation processes are possible in white dwarfs. In this work, we calculate the phase diagram of more complicated O-Ne-Fe mixtures and make predictions for the internal structure of the separated white dwarf. There are two possible outcomes determined by a complicated interplay between the Ne abundance, the $^{22}$Ne fraction, and the $^{56}$Fe abundance. Either Fe distills to form an inner core because the first O-Ne solids are buoyant, or an O-Ne inner core forms and Fe accumulates in the liquid until Fe distillation begins and forms a Fe shell. In the case of an Fe shell, a Rayleigh-Taylor instability may arise and overturn the core. In either case, Fe distillation may only produce a cooling delay of order 0.1 Gyr as these processes occur early at high white dwarf luminosities. Fe inner cores and shells may be detectable through asteroseismology and could enhance the yield of neutron rich elements such as $^{55}$Mn and $^{58}$Ni in supernovae. 
\end{abstract}



\section{Introduction}

In the absence of a heat source white dwarfs (WDs) cool monotonically, allowing the use of WD surface temperatures as a proxy for stellar ages. However, precision cosmochronology is only possible with accurate theories of WD cooling and detailed accounting of any heat sources which could delay cooling. A rapidly growing body of work studying plasma mixtures in WDs finds that the exact composition of the WD core impacts these heating processes. The composition of the first crystals that form is determined by the composition of the fluid in the core, as nuclei with higher charges experience greater Coulomb energies and thus crystallize at higher temperatures. We are therefore motivated in this work to study some of these mixtures in more detail.

The Gaia Data Release 2 (DR2) and Early Data Release 3 (EDR3) include 360,000 WD candidates \citep{gentile2019,gentile2021} and near-future work on the Gaia DR3 data may identify more and spectroscopically verify many. With such a large population of WDs, many internal processes that have long been theorized about have now been observationally confirmed. For example, \citealt{tremblay2019} finds a pile-up in the cooling sequence of WDs within 100 pc which is evidence for core crystallization. In addition, in late stellar burning CNO elements are converted in $^{22}$Ne, which experiences a net gravitational force in WD interiors and sinks toward the core, releasing gravitational potential energy as heat \citep{isern1991,bildsten2001gravitational}. $^{22}$Ne transport may be especially important to the so-called 'Q-branch' WDs which may be the result of WD mergers and have an observed multi-gigayear cooling delay \citep{cheng2019cooling,Bauer2020,camisassa2021,blouin2021}.

Strong separation processes can radically change the composition profile of a star, impacting observable astrophysics beyond cooling such as oscillation periods and supernova nucleosynthetic yields. Pulsation periods are sensitive to mean molecular weights, with \cite{Chidester2021} finding that the presence of absence of $^{22}$Ne may produce a systematic offset in relative period shifts of $\Delta P / P \approx \pm 0.5 \%$. However, this analysis could be considered agnostic to the exact isotope that shifts the mean molecular weight toward more neutron-rich values, and suggests that $^{56}$Fe separation could be important for future work modeling pulsation periods. Given the high precision (6 to 7 significant figures) of measured g-mode pulsation periods, it may be possible to infer some separation in WDs despite the small relative period shifts. 
Likewise, efficient separation resulting in compact inner cores with higher neutron excesses could impact type 1a nucleosynthesis. If the core reaches nuclear statistical equilibrium during detonation then the greater neutron fraction of the separated material may more easily produce neutron rich nuclides that would otherwise not be expected from rapidly burned symmetric nuclei. The observed abundances of $^{55}$Mn and $^{58}$Co suggest that type 1a supernovae are an important source of these nuclides \citep{seitenzahl2013solar,Yamaguchi2015}. Lastly, the accumulation of neutron rich nuclides decreases the core electron fraction, which increases the pressure and core electron Fermi energy. If the Fermi energy increases enough it may trigger electron capture reactions which could initiate a supernova in an isolated WD. Given that the WD must be very near the electron capture threshold for this to be possible such a channel would be rare, but such a delayed separation-induced supernova could be an interesting possibility for near-Chandrasekhar WD merger remnants that fail to explode initially \citep{caplan2020black,Caiazzo2021}.

Given the importance of separation processes to this range of astrophysical problems, theorists have now considered the separation of many nuclei rich nuclides. Separation of $^{22}$Ne in C-O WDs is now the most widely studied, with distillation seeming likely. Initial suggestions by \cite{Bauer2020} that $^{22}$Ne could directly separate from C-O and precipitate to the core were found to be incompatible with molecular dynamics simulations and the phase diagram of C-O-Ne by \cite{Caplan2020APJL}.
\cite{blouin2021} argues that the first crystals that form are likely a C-O alloy enhanced in O relative to the fluid but depleted in $^{22}$Ne. By excluding $^{22}$Ne these crystals become buoyant and a distillation process begins that may produce either a $^{22}$Ne inner core or a shell of $^{22}$Ne around a C-O-Ne inner core, depending on the exact $^{22}$Ne concentration. While $^{22}$Ne does not directly precipitate, nuclei with higher charges may separate more strongly, and \cite{caplan2021} shows that $^{56}$Fe may separate to form either a pure Fe solid or a C-O-Fe alloy of about 15\% Fe which precipitates to form an inner core in C-O WDs before $^{22}$Ne separates. The behavior of neutron rich nuclides in O-Ne WDs has also been studied. Much of the $^{22}$Ne gets converted into heavier nuclides such as $^{23}$Na, which \cite{blouin2021b} shows does not tend to separate from O-Ne mixtures due to the similar charges. 

Presently, Fe in O-Ne mixtures remains unstudied and so will be the subject of this work. These mixtures are more complicated because the $^{22}$Ne cannot phase separate on crystallization from the bulk $^{20}$Ne. Therefore, there is a complicated interplay impacting crystallization and the buoyancy of those crystals involving three parameters: the Ne abundance, the $^{22}$Ne concentration, and the $^{56}$Fe concentration. 

In Sec. \ref{sec:pd} we present a semi-analytic phase diagram of O-Ne-Fe mixtures. 
In Sec. \ref{sec:impact} we discuss the impacts of separation on the evolution of the WD and consider precipitation and distillation processes that may occur for varying concentrations of the neutron rich nuclides. We summarize in Sec. \ref{sec:conclusion}

\section{Phase diagrams}\label{sec:pd}

\begin{figure}
	\includegraphics[trim=0 0 0 
	0,clip,width=0.49\textwidth]{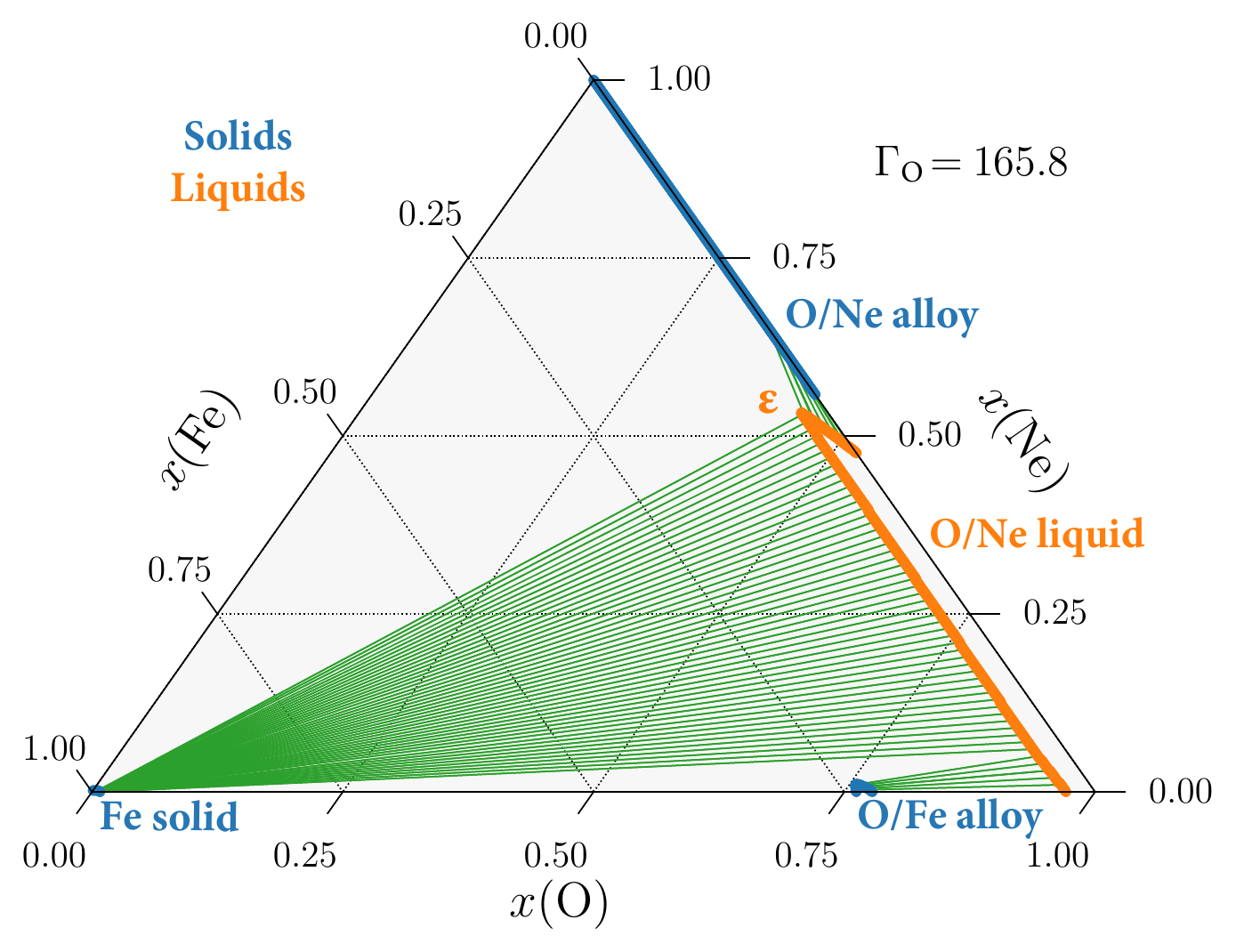}
    \caption{ONeFe phase diagram. Liquidus
(orange) and solidus (blue) curves are connected by tie-lines (green) showing coexisting solid and liquid compositions. To find a specific composition, lines of constant $x_i$ are projected from the slope of the tick marks. Along the right side axis one finds the O-Ne mixture with trace Fe, while the white region bounded by the liquidus is stable oxygen-rich liquid at the given temperature. The `corner' of the liquidus curve near $ (x_{\rm O}, x_{\rm Ne}, x_{\rm Fe}) = (0.44, 0.53, 0.03)$ labeled \textepsilon\ corresponds to the eutectic point where the liquid coexists with two solids, the pure Fe solid and the O-Ne solid that is depleted in Fe.}
    \label{fig:PD1}
\end{figure}

We begin by calculating a three-component phase diagram for O-Ne-Fe mixtures, shown in Fig. \ref{fig:PD1}, to understand what solid phases may coexist with a typical liquid composition present in a WD.

To summarize briefly, this semi-analytic method uses the analytic fits to the free energies found in \cite{ogata1993} to find co-existing solids and liquids that share a tangent plane on the minimum free energy surfaces. This method was developed by \cite{medin2010} and the approach for ternary phase diagrams is developed further in \cite{Caplan2018}. Our approach here is the same as in past work, which is now numerous in the literature \citep{Caplan2020APJL,blouin2021b,caplan2021}.

We present the phase diagram at $\Gamma_\mathrm{O} = 165.8$, which is approximately the condition for an equal mixture of O and Ne to begin freezing. As can be seen near the Ne axis, a range of mixtures with up to about $x_\mathrm{Fe} \approx 0.03$ freeze and form a O-Ne alloy which is enhanced in Ne relative to the liquid. This O-Ne alloy is also depleted in Fe. Liquids with increasing Fe coexist with increasingly Ne rich solids, but those Ne-rich solids are also depleted in Fe. Beyond $x_\mathrm{Fe} \approx 0.03$ we reach the eutectic point (\textepsilon), where the liquid can coexist with a pure solid. Given the high $x_\mathrm{Fe}$ of the eutectic point, this suggests that direct Fe precipitation from the liquid, like was found for C-O-Fe mixtures in \cite{caplan2021}, is unlikely for O-Ne-Fe mixtures. 

Other features of the phase diagram are interesting but less astrophysically relevant. For example, near the bottom-right corner of the diagram one finds very O rich mixtures. For liquids above about $x_\mathrm{O} \approx 0.85$, we find coexistence with a Ne depleted solid that is weakly enhanced in Fe. While the ratio of O to Ne in a WD remains uncertain, such a low ratio of Ne/O seems unlikely and will not be considered further in this work. 

The behavior of the phase diagram at other $\Gamma_\mathrm{O}$ is similar to this case. As $\Gamma_{\mathrm{O}}$ increases the eutectic point moves toward the bottom-right corner as mixtures with lower average charge begin freezing. The liquidus curve also approaches the Ne axis, as fluids with lower Fe begin to coexist with the pure Fe solid. The last mixtures to freeze appear near the pure O corner at $\Gamma_\mathrm{O}\approx 185$, and has roughly $x_\mathrm{Fe} \approx 0.01$, as a small impurity concentration is known to lower the melting temperature (see e.g. \citealt{ogata1993,medin2010}). Thus, as the eutectic point moves upward toward increasingly Ne rich mixtures, the fluid can only coexist with the pure Fe solid when the fluid is significantly enhanced in Fe.

The behavior of the phase diagram at very low Fe abundance determines what solid phase forms first, so it must be understood to determine the evolution of the liquid composition. The charge ratios in these mixtures are large ($Z_{\mathrm{Fe}} / Z_{\mathrm{O}}=3.25$), and the abundance of Fe in the O-Ne at solar composition is less than one percent. As a consequence, the free energies from \cite{ogata1993} that are used to calculate the phase diagram are extrapolated to lower concentrations than they were fit (the lowest number fractions explored in \cite{ogata1993} are 0.01 for $Z_i/Z_j = 3$). Preliminary work studying mixtures of O-Ne with trace Fe using molecular dynamics simulations are ongoing and will be presented in future work. 

\section{Impact on the evolution of ultramassive white dwarfs}\label{sec:impact}
\subsection{Do the Fe-depleted crystals float?}
We have established in Section~\ref{sec:pd} that Fe-depleted solids will form when an O-Ne-Fe mixture with a composition typical of ultramassive O-Ne white dwarfs freezes. Because of $^{56}$Fe's extra neutrons ($Z=26$, $A=56$), the Fe-depleted crystals could be lighter than the coexisting liquid, which would make them buoyant and trigger a distillation process analogous to that predicted for $^{22}$Ne in C-O white dwarfs \citep{isern1991,segretain1996,blouin2021}. Whether the solid phase has a lower density than the liquid phase will depend on the exact composition of the initial liquid mixture, but mostly on the abundances of the neutron-rich isotopes $^{56}$Fe and $^{22}$Ne. As $X(^{56}{\rm Fe})=0$ in the solid phase, a higher $X(^{56}{\rm Fe})$ in the liquid will favor distillation since it will maximize the $^{56}$Fe abundance gradient between both phases.\footnote{Note that these are mass fractions rather than the number fractions denoted by $x$ in the earlier sections.} In contrast, a higher $X(^{22}{\rm Ne})$ tends to inhibit distillation. When an O-Ne mixture freezes, the solid phase is enriched in Ne (Figure~\ref{fig:PD1}, \citealt{camisassa2019,blouin2021b}). This separation affects all Ne isotopes equally: the $^{22}$Ne/$^{20}$Ne ratio in the solid phase remains the same as in the liquid phase.\footnote{Gravitational settling of neutron-rich isotopes in the liquid phase is negligible in O-Ne ultramassive white dwarfs given how early they crystallize \citep{schwab2021}.} This means that the solid phase will be enriched in $^{22}$Ne, thereby increasing its density.

\begin{figure}
	    \includegraphics[width=0.99\columnwidth]{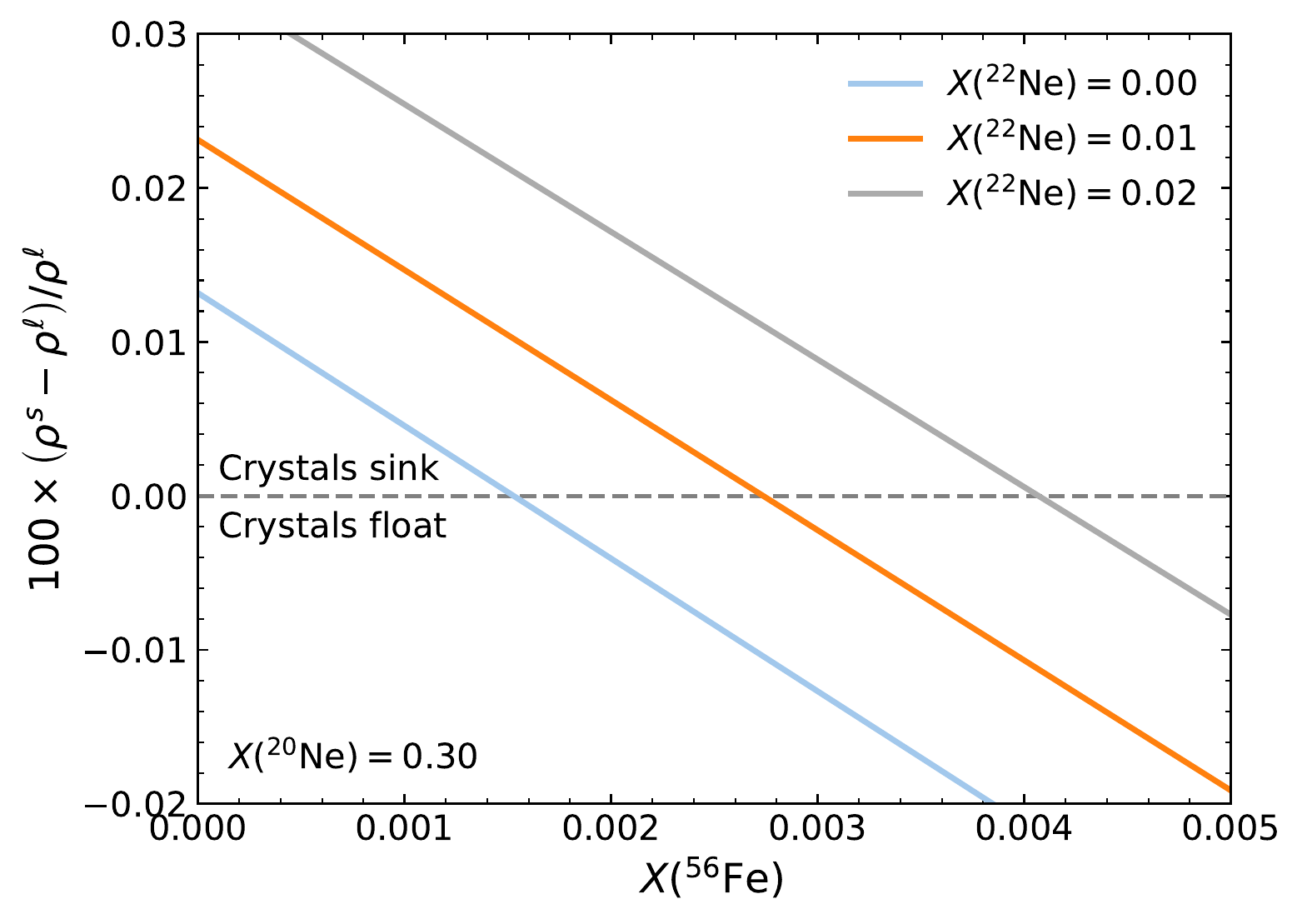}
    \caption{Difference in mass density between the coexisting solid and liquid phases as a function of the $^{56}$Fe mass fraction in the liquid phase. The three lines correspond to three different $^{22}$Ne abundances in the liquid, as indicated in the legend. $X(^{20} {\rm Ne})$ is fixed to 0.30 and $X(^{16}{\rm O}) = 1-X(^{20}{\rm Ne}) - X(^{22}{\rm Ne}) - X(^{56}{\rm Fe})$.}
    \label{fig:density_change}
\end{figure}

Figure~\ref{fig:density_change} illustrates these two competing effects, assuming a typical $^{20}$Ne mass fraction $X(^{20}{\rm Ne})=0.30$ and $X(^{16}{\rm O}) = 1-X(^{20}{\rm Ne}) - X(^{22}{\rm Ne}) - X(^{56}{\rm Fe})$. Each line indicates how the density difference between the liquid and solid varies as a function of the $^{56}$Fe abundance in the liquid and for different $^{22}$Ne abundances. Without any $^{22}$Ne, $X(^{56}{\rm Fe}) \gtrsim 0.0015$ is enough to make the solid buoyant ($\rho^{s} < \rho^{\ell}$) and trigger distillation.\footnote{$0 < X(^{56}{\rm Fe}) < 0.0015$ is not sufficient to trigger distillation as the $^{56}$Fe depletion of the solid phase is then too small to counteract the increased density of the solid phase compared to the liquid phase.} With a higher $^{22}$Ne content, more $^{56}$Fe is needed as the enrichment of $^{22}$Ne in the solid phase needs to be counterbalanced by a stronger $^{56}$Fe depletion. The liquid and solid densities used to generate Figure~\ref{fig:density_change} were calculated using the same multi-component ionic mixture model as \citet[which relies on the free energy analytical fits of \citealt{dubin1990,ogata1993,dewitt2003}]{medin2010}, to which we added the contribution of the degenerate electron gas. The liquid and solid densities were evaluated at a fixed temperature $T=1.7 \times 10^7\,{\rm K}$ and pressure $P= 4 \times 10^{25}\,{\rm erg}\,{\rm cm}^{-3}$, typical values for the central layers of an ultramassive white dwarf at the onset of crystallization.\footnote{For comparison, $T \simeq 4 \times 10^6\,{\rm K}$ and $P \simeq 2 \times 10^{23}\,{\rm erg}\,{\rm cm}^{-3}$ at the center of a 0.6\,$M_{\odot}$ white dwarf when crystallization of the C-O core starts.}

\begin{figure*}
\centering
	    \includegraphics[width=1.5\columnwidth]{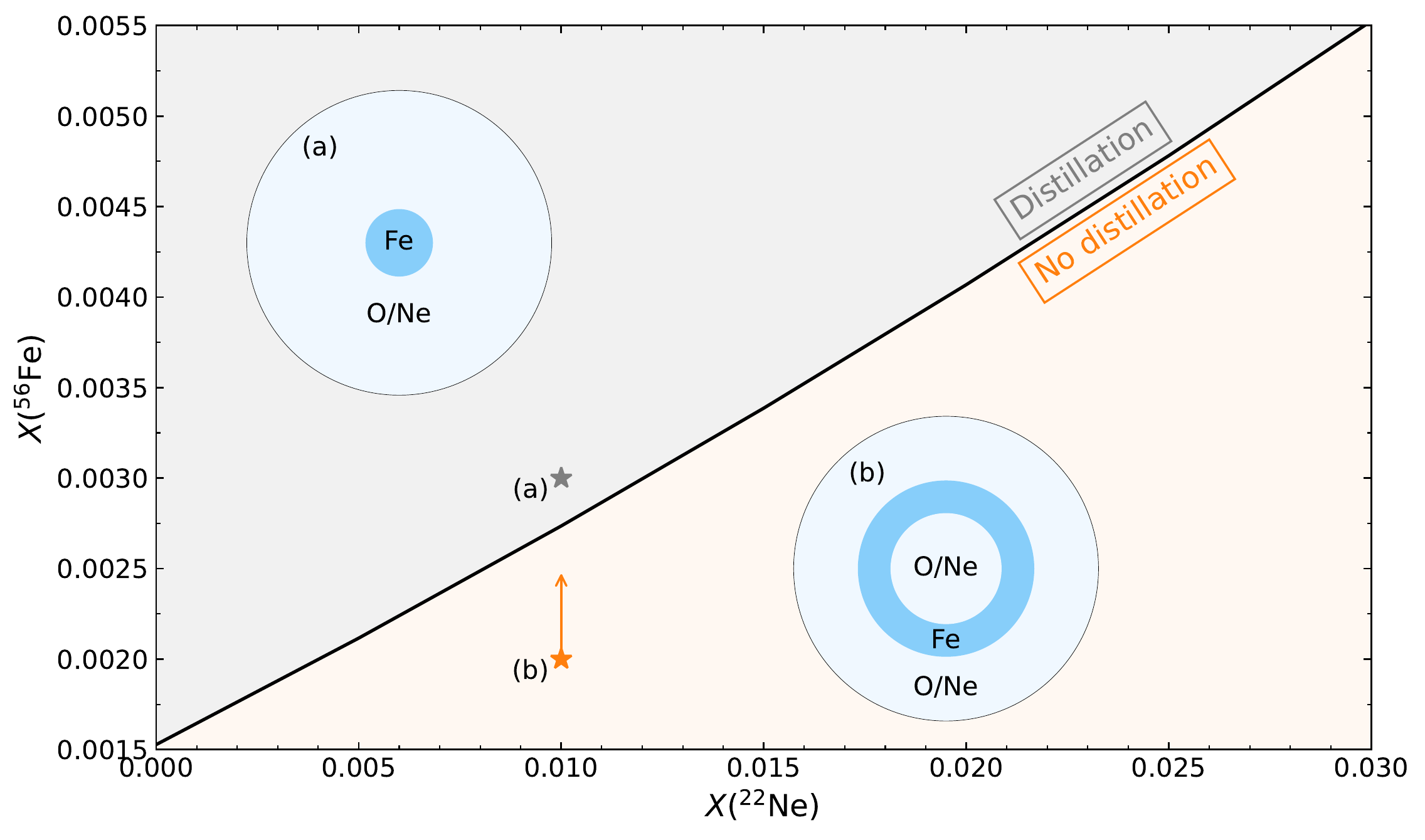}
    \caption{The black line indicates composition where the solid phase has neutral buoyancy ($\rho^{s}=\rho^{\ell}$). In the grey region above this boundary, the $^{56}$Fe depletion in the solid phase makes the crystal lighter than the liquid, leading to distillation. In the orange region below, the change in $^{56}$Fe abundance between both phases is too small to make the crystal buoyant. As in Figure~\ref{fig:density_change}, $X(^{20} {\rm Ne})$ is fixed to 0.30 and $X(^{16}{\rm O}) = 1-X(^{20}{\rm Ne}) - X(^{22}{\rm Ne}) - X(^{56}{\rm Fe})$. Two scenarios are illustrated. In scenario (a), a $^{56}$Fe-rich white dwarf forms a Fe central core through distillation. In scenario (b), the white dwarf initially forms a solid Fe-depleted core. As this central solid core is formed, the $^{56}$Fe abundance in the liquid increases (orange arrow). This enrichment ultimately allows distillation to take place, leading to the formation of an Fe shell.}
    \label{fig:distillation_diagram}
\end{figure*}

Figure~\ref{fig:distillation_diagram} recasts the results of Figure~\ref{fig:density_change} to show which $^{56}$Fe and $^{22}$Ne abundances lead to the formation of buoyant solids (using the same assumptions on $X(^{20}{\rm Ne})$ and $X(^{16}{\rm O})$ as in Figure~\ref{fig:density_change}). The upper grey region contains compositions where the crystals will float and where distillation is expected to take place; in the lower orange region, the crystals remain heavier than the liquid. As we will see, this dichotomy implies two different outcomes for crystallizing white dwarfs. Note that the exact position of the boundary shown in Figure~\ref{fig:distillation_diagram} should not be taken too literally. In particular, we have neglected the effect of other neutron-rich species ($^{23}$Na, $^{25}$Mg) that could impact the density of the solid phase. 

\subsection{Scenario 1: Formation of an Fe central core}
Let us first consider a scenario in which there is enough $^{56}$Fe to trigger distillation at the onset of crystallisation. Neglecting the effect of other minor species, Figure~\ref{fig:distillation_diagram} shows that this would require $X(^{56}{\rm Fe}) \gtrsim 0.0015 + 0.14 X(^{22}{\rm Ne})$. This scenario appears unlikely for an O-Ne white dwarf formed through single-star evolution as in that case we expect $X(^{56}{\rm Fe}) \simeq 0.1 X(^{22}{\rm Ne})$ ($X(^{22}{\rm Ne}) \simeq Z$ and $X(^{56}{\rm Fe}) \simeq 0.1 Z$, where $Z$ is the progenitor metallicity). In other words, the high initial metallicity required to attain a large $^{56}$Fe abundance would also increase the $^{22}$Ne abundance, thereby preventing the formation of buoyant crystals. However, this scenario remains plausible for a white dwarf formed from the merger of two C-O white dwarfs. In that case, the $^{22}$Ne abundance is negligible (\citealt{schwab2021} give $X(^{22}{\rm Ne})=0.0022$) and $X(^{56}{\rm Fe}) \gtrsim 0.0018$ would then be sufficient to trigger distillation. The $^{56}$Fe abundance threshold may even be lower than this given the relatively large $^{25}$Mg abundance predicted in those objects (\citealt{schwab2021} give $X(^{25}{\rm Mg})=0.015$). \cite{blouin2021b} have shown that in an O-Ne-Mg plasma, Mg is depleted from the solid phase during crystallization. This Mg depletion has no important effect of the buoyancy of the crystals when the $^{25}$Mg mass fraction is negligible and most of the Mg is under the $^{24}$Mg form (as is the case for white dwarfs that have evolved through a single-star evolutionary channel), but it will contribute to make the crystals lighter if there is a substantial amount of $^{25}$Mg. Therefore, in the case of O-Ne white dwarfs formed through a merger, we expect that Mg depletion in the crystals will complement $^{56}$Fe separation and render the crystals buoyant at $^{56}$Fe abundances that are lower than the threshold given by Figure~\ref{fig:distillation_diagram}. Note that a similar effect is not expected from $^{23}$Na separation \citep[][Figure~6]{blouin2021b}.

Assuming that the crystals formed at the onset of crystallization in the central layers are buoyant, how will the composition profile of the star change due to fractionation? As in the case of $^{22}$Ne distillation, the $^{56}$Fe abundance at the center of the star gradually increases. Buoyant $^{56}$Fe-depleted crystals float up and displace $^{56}$Fe-rich liquid toward the center. This increase in the central $^{56}$Fe abundance will inevitably push the liquid mixture to the critical (eutectic) composition where the liquidus has a sharp kink (point \textepsilon\ in Figure~\ref{fig:PD1}). After reaching this point, crystals that are strongly enriched in $^{56}$Fe are produced and a Fe-rich central core is gradually formed. Note that distillation continues to operate just above the crystallization front: this is how $^{56}$Fe from the outer layers is transported to the central region. This is somewhat different from $^{22}$Ne distillation in C-O white dwarfs, where the distillation of a given layer ends when an azeotropic point is reached and the mixture freezes at a constant composition. Here, the compositions of the solid and liquid phases are never the same.

Before evaluating the impact of this process on white dwarf cooling, we need to establish what fraction of the total $^{56}$Fe content of the star will be transported to the center. As the process described in the previous paragraph unfolds, the $^{56}$Fe abundance in the liquid above the distilling region will decrease and eventually reach the point where neutrally buoyant crystals would be formed (black line in Figure~\ref{fig:distillation_diagram}). Does distillation stop there? If neutrally buoyant Fe-depleted crystals are formed, then the Fe abundance in the liquid will immediately increase and the composition should again be such that buoyant crystals are formed. The formation of neutrally buoyant (or sinking) crystals therefore appears to be an unstable state: the composition will presumably always return to the distillation regime. From this consideration, we speculate that the distillation process will not stop until all the Fe content of the star is distilled and transported to the central layers. Note that the same argument can be made for $^{22}$Ne distillation in C-O white dwarfs.

To calculate the energy released by the formation of the Fe central core, we calculate static white dwarf structures using STELUM (\citealt{bedard2022} and references therein). Two structures are calculated: one where we assume an homogeneous distribution of Fe in the core and one where all the Fe is concentrated at the center. The difference in binding energy between these two states,
\begin{equation}
    \Delta B = - \left( \Delta \Omega + \Delta U \right),
\end{equation}
where $\Delta \Omega$ is the change of the gravitational energy and $\Delta U$ the change of the internal energy, corresponds to the energy released during the distillation of $^{56}$Fe. Table~\ref{tab:deltaB} gives values of $\Delta B$ for two different stellar masses and two different $^{56}$Fe abundances. The energy released is one order of magnitude smaller than the energy released by the formation of a $^{22}$Ne-rich central core in ultramassive C-O white dwarfs \citep{blouin2021}, which simply reflects the fact that there is roughly ten times less $^{56}$Fe than $^{22}$Ne.

\begin{table}
	\centering
	\caption{Energy released during the formation of an Fe central core.}
	\label{tab:deltaB}
	\begin{tabular}{ccccc} 
		\hline
		$M/M_{\odot}$ & $X(^{56}{\rm Fe})$ & $L/L_{\odot}$$^a$ & $\Delta B$ ($10^{47}$ erg) & $\Delta \tau$ (Gyr) \\
		\hline
                1.05 & 0.002 & -2.2 & 0.8 & 0.10 \\
                1.05 & 0.004 & -2.2 & 1.5 & 0.20 \\
                1.25 & 0.002 & -1.5 & 1.8 & 0.05 \\
                1.25 & 0.004 & -1.5 & 3.5 & 0.10 \\
        \hline
        \multicolumn{5}{p{6cm}}{$^a$ Luminosity at the onset of crystallization}
		\end{tabular}
\end{table}

We can estimate the cooling delay $\Delta \tau$ resulting from the transport of $^{56}$Fe by dividing $\Delta B$ by the luminosity $L$ at which crystallization starts. In reality, the heating will occur over a range of luminosities (this is not an instantaneous process), but distillation proceeds rapidly enough that assuming a constant $L$ is a reasonable approximation (but will slightly underestimate the actual cooling delay). The resulting cooling delays are given in Table~\ref{tab:deltaB}. A priori, one could expect longer cooling delays than in the case of $^{56}$Fe precipitation in C-O cores \citep{caplan2021}, as here all the $^{56}$Fe is transported to the center, contrary to $^{56}$Fe precipitation in C-O. However, we find comparable cooling delays (see \citealt{salaris2022} for estimates of the cooling delay due to $^{56}$Fe precipitation in C-O white dwarfs). This is explained by the fact that the larger energy release is counterbalanced by a larger luminosity. The crystallization of an O-Ne core starts earlier on the white dwarf cooling track than the crystallization of a C-O core, a direct consequence of the higher ionic charges of the O-Ne mixture. For that reason, this process is unable to generate cooling delays larger than $\simeq 0.2\,$Gyr, even when assuming a very large $^{56}$Fe abundance ($X(^{56}{\rm Fe})=0.004$ in Table~\ref{tab:deltaB} corresponds to a progenitor metallicity of $Z \simeq 0.04$). 

\added{While the 0.05-0.20 Gyr cooling delays of Table~\ref{tab:deltaB} are small over the multi-Gyr evolution of white dwarfs, they are quite sizeable when compared to the age of the white dwarf when the Fe phase separation takes place. Although \cite{camisassa2019} does not report on the exact masses we consider here, their Figure 5 does show the luminosity as a function of cooling time for a range of white dwarf masses between 1.10 $M_\odot$ and 1.29 $M_\odot$. One could estimate from this that Fe phase separation begins at a cooling age of ~0.4 Gyr in the 1.05 $M_\odot$ case and ~0.2 Gyr in the 1.25 $M_\odot$ case, suggesting that the cooling delay can be comparable to the age of the white dwarf when the Fe phase separation takes place. For massive white dwarfs currently undergoing Fe phase separation, one might also expect them to lean toward super-solar metallicities and iron abundances as their progenitors would have formed more recently than the sun.}

\subsection{Scenario 2: Formation of an Fe shell}
As explained in the previous section, a core composition that lies below the neutral buoyancy boundary of Figure~\ref{fig:distillation_diagram} is expected for most O-Ne white dwarfs. In that case, Fe-depleted crystals are still formed, but they do not float up and no distillation can take place. As the crystallization front progresses outward and Fe is expelled from each new solid layer, $X(^{56}{\rm Fe})$ in the liquid increases (orange arrow in Figure~\ref{fig:distillation_diagram}). Inevitably, the Fe abundance will eventually reach the neutral buoyancy line and distillation will ensue. A Fe-rich shell will then be formed around the Fe-depleted solid central core. Analogous to the previous scenario, all the Fe content of the remaining liquid layers will be transported to this Fe shell. This process will release a fraction of the energy generated by the formation of an Fe central core.

To first order, this Fe shell will form when the solid Fe-depleted core reaches a mass of
\begin{equation}
    M_{\rm solid} = M_{\star} \left( 1- \frac{X_0 (^{56}{\rm Fe})}{X_{\rm c} (^{56}{\rm Fe})} \right),
    \label{eq:msolid}
\end{equation}
where $X_0 (^{56}{\rm Fe})$ is the initial Fe mass fraction (assumed to be constant throughout the core) and $X_{\rm c} (^{56}{\rm Fe})$ is the minimum Fe abundance required to trigger distillation. Note that this estimate does not take into account two potentially important effects. Firstly, as the solid Fe-depleted core is formed, the Ne abundance in the liquid will change, thereby changing the value of $X_{\rm c} (^{56}{\rm Fe})$. Secondly, Equation\,\ref{eq:msolid} assumes that the $^{56}$Fe added to the liquid upon the crystallisation of each new layer is homogeneously mixed throughout the liquid portion of the core. For this assumption to hold, the density of the liquid layer immediately above the crystallization front must become lower than the density of the next liquid layer when the solid core grows by some amount. This density inversion then triggers mixing through a Rayleigh--Taylor instability, and an homogeneous liquid core composition can be achieved \citep{isern1997,camisassa2019}. This density inversion is achieved through the decrease of the Ne abundance in the liquid layer just above the crystallization front. However, the enrichment of $^{56}$Fe in the liquid phase competes with this mechanism. Therefore, if $^{56}$Fe fractionation is strong enough (which necessitates high $^{56}$Fe abundances), convective mixing can stop before the neutral buoyancy threshold is reached. In that case, $X(^{56} {\rm Fe})$ in the liquid just above the crystallization front will increase rapidly and distillation will start earlier than predicted by Equation~\ref{eq:msolid}.

\section{Conclusion}\label{sec:conclusion}

Despite the complexity of O-Ne-Fe mixtures in white dwarfs, we find two coarse outcomes: either Fe distills to form an inner core because the first O-Ne solids are buoyant, or an O-Ne inner core forms and Fe accumulates in the liquid until Fe distillation begins and forms a Fe shell. 

It is an open question whether the Fe shell is hydrodynamically stable. 
Pressure continuity at the interface between the Fe shell and the O-Ne core implies that there is a density inversion that may be subject to a Rayleigh-Taylor instability. With lower $Y_e$ the equation of state softens, and so the density is greater on the Fe side of the interface ($Y_e^{\rm Fe} = 0.464$) than in the O-Ne core ($Y_e^{\rm O-Ne} \approx 0.50$) resulting in a density discontinuity of order $\Delta \rho / \rho \approx 0.077$. Such a density inversion implies that the O-Ne crystal is buoyant relative to the Fe shell, possibly triggering an overturn once the Fe shell reaches some critical size. The Rayleigh-Taylor instability at solid-solid interfaces in degenerate plasmas is a complicated problem but has been considered in past work. In a simplified model for neutron star crusts, \cite{blaes1990} calculates a maximum density discontinuity that can be supported by relativistic degenerate electrons in an ionic bcc lattice to be approximately $\Delta \rho / \rho \approx 0.078 (Z/26)^{(2/3)}$, suggesting that overturn is possible. 

Future work is required to understand this overturn process. One possibility is that a macroscopic shell must form before the buoyant instability overcomes the mechanical strength of the crystal, at which point there is a rapid catastrophic overturn of the core. Another possibility is that small Fe crystallites form and grow to some critical mesoscopic size, after which they begin to migrate through the O-Ne crystal toward the center of the star. These possibilities motivate a detailed study of Coulomb crystals with mixtures, and especially diffusion coefficients in O-Ne-Fe mixtures due to their importance for viscoelastic creep. In either case, the transport of the entire mass of the Fe shell to the center of the star will result in a similar heat release as the direct formation scenario. Depending on the critical density discontinuity required to trigger overturn, this process may also be relevant to $^{22}$Ne-rich solid shells formed in C-O white dwarfs.

Given the robustness of these separation processes, we motivate the inclusion of separation processes in many other models. Asteroseismologists should consider Fe separation and the presence of Fe inner cores (or shells) when modeling g-mode oscillations, in addition to $^{22}$Ne. Supernova simulations should likewise test the effects of Fe cores and shells to determine their effect on the explosion, especially nucloesynthetic yields of neutron-rich nuclides such as $^{55}$Mn and $^{58}$Ni. Depending on the delay time for mergers, Fe inner cores could form which may impact the detonation and could be studied with hydrodynamic simulations of WD mergers. As with all separation processes, diffusion coefficients impact settling rates but they can be readily determined from molecular dynamics simulations. While \cite{caplan2022accurate} finds a general law for diffusion in mixtures it has not been checked for charge ratios as large as C-O-Fe and O-Ne-Fe, which should be investigated in future work.

\begin{acknowledgments}
We thank Mar\'ia Camisassa for helpful discussions. This work was supported in part by the National Science Foundation under Grant No. OISE-1927130 (IReNA). SB is a Banting Postdoctoral Fellow and a CITA National Fellow, supported by the Natural Sciences and Engineering Research Council of Canada (NSERC). M.C. and S.B. thank the KITP for hospitality and M.C. acknowledges support as a KITP Scholar. This research was supported in part by the National Science Foundation under Grant No. NSF PHY-1748958. 
This work benefited from support by the National Science Foundation under Grant No. PHY-1430152 (JINA Center for the Evolution of the Elements).
\end{acknowledgments}

\bibliographystyle{aasjournal}



\end{document}